\documentclass[english,reprint,tightenlines,eqsecnum,aps,prd,nofootinbib]{revtex4}
\usepackage[T1]{fontenc}
\usepackage[latin9]{inputenc}
\usepackage{amsmath}

\makeatletter
\@ifundefined{textcolor}{}
{%
 \definecolor{BLACK}{gray}{0}
 \definecolor{WHITE}{gray}{1}
 \definecolor{RED}{rgb}{1,0,0}
 \definecolor{GREEN}{rgb}{0,1,0}
 \definecolor{BLUE}{rgb}{0,0,1}
 \definecolor{CYAN}{cmyk}{1,0,0,0}
 \definecolor{MAGENTA}{cmyk}{0,1,0,0}
 \definecolor{YELLOW}{cmyk}{0,0,1,0}
}

\makeatother

\usepackage{babel}
\begin{document}

\title{Late-time decay of coupled electromagnetic and gravitational perturbations
outside an extremal charged black hole}

\author{Orr Sela}

\address{Department of physics, Technion-Israel Institute of Technology, Haifa
32000, Israel}
\begin{abstract}
In this paper we employ the results of a previous paper on the late-time
decay of scalar-field perturbations of an extreme Reissner-Nordstrom
black hole, in order to find the late-time decay of coupled electromagnetic
and gravitational perturbations of this black hole. We explicitly
write the late-time tails of Moncrief's gauge invariant variables
and of the perturbations of the metric tensor and the electromagnetic
field tensor in the Regge-Wheeler gauge. We discuss some of the consequences
of the results and relations to previous works.
\end{abstract}
\maketitle

\section{Introduction}

Coupled electromagnetic and gravitational perturbations of the Reissner-Nordstrom
black hole have been studied by various authors, and several formalisms
were developed for studying them {[}1-11{]}. In Refs. {[}2-4{]}, Moncrief
found, using the Hamiltonian formulation of the Einstein-Maxwell equations,
gauge invariant variables (under both electromagnetic gauge transformations
and infinitesimal diffeomorphisms) which can be used as the perturbation
variables. In particular, one can use these gauge invariant variables
in order to express all metric and electromagnetic perturbations after
specifying the gauge. In Ref. \cite{Bicak 79}, Bicak related the
various formalisms and explicitly wrote all metric and electromagnetic
perturbations in terms of Moncrief's gauge invariant variables in
the Regge-Wheeler gauge. 

Later, in \cite{Bicak 80}, Bicak employed his results from \cite{Bicak 79}
and showed that scalar-field perturbations serve as a prototype for
the coupled electromagnetic and gravitational perturbations; he then
used his results from Ref. \cite{Bicak 72} that analyzed scalar-field
perturbations to deduce the late-time behavior of the coupled perturbations.
In both cases (scalar and coupled perturbations), Bicak obtained a
late-time asymptotic behavior that corresponds to a power law decay
\footnote{In this statement, we exclude perturbations that correspond to a slowly
rotating Kerr-Newman black hole.%
}. This behavior of asymptotic power law decay of perturbations also
appears in the simpler case of a Schwarzschild black hole (see, for
example, Refs. {[}14-15{]}). 

In a recent paper \cite{Sela}, we revisited Bicak's analysis \cite{Bicak 72}
of the scalar-field perturbations, concentrating on the \emph{extremal}
Reissner-Nordstrom (ERN) black hole. In this case, we found that the
late-time asymptotic behavior of the scalar perturbations is again
of the form of a power law decay, but with a different exponent compared
with the one obtained by Bicak (for the same form of initial data).
The power law that we found is exactly the same as the one obtained
in \cite{Lucietti} using both numerical computations (that correspond
to some specific cases) and analytical derivations. 

In this paper, we use the results of \cite{Sela} to find the late-time
decay of coupled electromagnetic and gravitational perturbations in
ERN geometry, in complete analogy with the derivations of Bicak's
\cite{Bicak 80}. In other words, we use our understanding of the
scalar-field late-time decay from \cite{Sela}, and the fact that
scalar-field perturbations serve as a prototype for the coupled perturbations
\cite{Bicak 80}, to find the late-time decay of all the relevant
quantities describing coupled perturbations in ERN spacetime. Moreover,
we examine how this decay changes under different choices of initial
data. In this, we revisit the analysis of \cite{Bicak 80} and develop
it further. 

The organization of this paper is as follows. In Sec. II, we introduce
the decoupled wave equations satisfied by certain combinations $\Psi_{\pm}$
of Moncrief's gauge invariant variables, and the close resemblance
they have to the scalar-field wave equation. We then deduce, as in
\cite{Bicak 80}, that scalar-field perturbations serve as a prototype
for coupled perturbations. In Sec. III, we describe the various kinds
of initial data we can consider for the perturbations, and classify
them. Later, in Sec. IV, we use the results of the previous paper
\cite{Sela} to find the late-time tails of $\Psi_{\pm}$ corresponding
to different choices of initial data. We then readily determine the
late-time tails of Moncrief's gauge invariant variables in Sec. V
using the definitions of the combinations $\Psi_{\pm}$. In Sec. VI,
we employ relations from Ref. \cite{Bicak 79} and find the late-time
decay of the perturbations of the metric tensor and the electromagnetic
field tensor in the Regge-Wheeler gauge. We conclude in Sec. VII.

\section{scalar-field perturbations as a prototype for coupled perturbations}

The ERN geometry is given in Schwarzschild-like coordinates by the
line element
\[
ds^{2}=-\left(1-M/r\right)^{2}dt^{2}+\left(1-M/r\right)^{-2}dr^{2}+r^{2}d\Omega^{2},
\]
where $M$ is the mass of the black hole and $d\Omega^{2}=d\theta^{2}+\sin^{2}\left(\theta\right)d\phi^{2}$.
In this paper, we focus on the external domain, $r>M$. Throughout
the analysis, we use the ``tortoise coordinate'', $r_{*}\left(r\right)$,
defined in the usual way by $dr/dr_{*}=\left(1-M/r\right)^{2}$. Fixing
the integration constant by setting $r_{*}\left(2M\right)=0$, we
get 
\[
r_{*}\left(r\right)=r-M-\frac{M^{2}}{r-M}+2M\ln\left(\frac{r}{M}-1\right).
\]
This function diverges to $+\infty$ at $r\rightarrow\infty$, to
$-\infty$ at $r\rightarrow M$, and vanishes at $r=2M$. At the asymptotic
regions $r\rightarrow\infty$ and $r\rightarrow M$, we can find the
inverse function $r\left(r_{*}\right)$ iteratively, 
\begin{equation}
r\sim r_{*}-2M\ln\left(\frac{r_{*}}{M}\right)\quad,\quad r_{*}\rightarrow\infty,\label{eq: r(s) inf}
\end{equation}
and 
\begin{equation}
r\sim M+\frac{M^{2}}{\left|r_{*}\right|}\left[1+\frac{2M}{\left|r_{*}\right|}\ln\left(\frac{\left|r_{*}\right|}{M}\right)\right]\quad,\quad r_{*}\rightarrow-\infty.\label{eq: r(s) horizon}
\end{equation}

When considering coupled electromagnetic and gravitational perturbations
of ERN black hole, one exploits the spherical symmetry of the background
and expands the perturbations in scalar, vector, and tensor harmonics.
Then, one identifies the corresponding parity (under inversion transformation)
and distinguishes between even and odd parity perturbations. This
way, for each $l$ in the harmonic expansion, we have even and odd
parity perturbations (and the corresponding perturbation equations
for the two types of parity decouple) {[}1-11{]}. 

In Moncrief's gauge invariant formalism {[}2-4{]}, the $l=1$ perturbations
are fully determined by a gauge invariant function $P_{f}\left(r,t\right)$
in the odd parity case and by a gauge invariant function $H\left(r,t\right)$
in the even parity case. For $l\geq2$, the odd parity perturbations
are determined by two gauge invariant functions, $\hat{\pi}_{f}\left(r,t\right)$
and $\hat{\pi}_{g}\left(r,t\right)$, and the even parity perturbations
also by two gauge invariant functions, $H\left(r,t\right)$ and $Q\left(r,t\right)$.%
\footnote{Throughout the analysis, our notations of the various functions are
similar to those used by Bicak in \cite{Bicak 79,Bicak 80}, and are
very close to the notations used by Moncrief in {[}2-4{]}.%
} 

Introducing the standard combinations of Moncrief's gauge invariant
functions (for $l\geq2$) {[}2-4,9-11{]},
\begin{equation}
P_{\pm}=\left(2\sigma\right)^{-1/2}\left[\pm\left(\sigma\pm3M\right)^{1/2}\hat{\pi}_{f}+\left(\sigma\mp3M\right)^{1/2}\hat{\pi}_{g}\right]\label{P+-}
\end{equation}
and 
\begin{equation}
R_{\pm}=\left(2\sigma\right)^{-1/2}\left[\left(\sigma\pm3M\right)^{1/2}H\mp\left(\sigma\mp3M\right)^{1/2}Q\right],\label{R+-}
\end{equation}
where (in the extremal case)
\begin{equation}
\sigma=M\left(2l+1\right),\label{eq: sig}
\end{equation}
we use the conventional notations $\Psi_{\pm}$ to denote $P_{\pm}$
in the case of odd perturbations and $R_{\pm}$ in the case of even
perturbations. For $l=1$, $\Psi_{+}$ denotes $P_{f}$ and $H$ for
odd and even perturbations, respectively. $\Psi_{-}$ has no meaning
for $l=1$. 

Now, we can describe the dynamics of all the coupled electromagnetic
and gravitational perturbations of an ERN black hole by the following
set of two decoupled wave equations satisfied by $\Psi_{\pm}$ \cite{Bicak 79,Bicak 80}:
\begin{equation}
\Psi_{\pm,tt}-\Psi_{\pm,r_{*}r_{*}}+V_{l\pm}^{\mathrm{odd,even}}\left(r_{*}\right)\Psi_{\pm}=0,\label{eq: wave eq.}
\end{equation}
where the effective potentials $V_{l\pm}^{\mathrm{odd,even}}$ are
given in the extremal case (ERN) by \cite{Bicak 79,Bicak 80}
\[
V_{l\pm}^{\mathrm{odd}}=\frac{1}{r^{2}}\left(1-\frac{M}{r}\right)^{2}\left(L-\frac{3M}{r}+\frac{4M^{2}}{r^{2}}\pm\frac{\sigma}{r}\right)
\]
and
\[
V_{l\pm}^{\mathrm{even}}=\left(1-\frac{M}{r}\right)^{2}\left(V\pm\sigma S\right),
\]
where
\[
S=\frac{1}{\left(r\varLambda\right)^{2}}\left[\frac{L^{2}-4}{r}+\frac{12M}{r^{2}}\left(1-\frac{M}{r}+\frac{M^{2}}{3r^{2}}\right)\right],
\]
\[
V=\frac{1}{\left(r\varLambda\right)^{2}}\left\{ \left(L-2\right)\left[L\left(L-2\right)+3\left(3L-2\right)\frac{M}{r}-\frac{4M^{2}}{r^{2}}\left(L-4+\frac{16M}{r}-\frac{6M^{2}}{r^{2}}\right)\right]\right.
\]
\[
\left.+\frac{4}{r^{2}}\left[9M^{2}\left(L-1+\frac{M}{r}\right)-\frac{M^{2}}{r^{2}}\left(\frac{8M^{4}}{r^{2}}-\frac{32M^{3}}{r}+39M^{2}\right)\right]\right\} ,
\]
\[
L=l\left(l+1\right)\quad,\quad\varLambda=L-2\left(1-\frac{M}{r}\right)\left(1-\frac{2M}{r}\right),
\]
and $\sigma$ is given by Eq. \eqref{eq: sig}. Note that for $l=1$,
Eq. \eqref{eq: wave eq.} is meaningful only for $\Psi_{+}$. As mentioned
above, all the electromagnetic and gravitational perturbations can
be obtained once all the $\Psi_{\pm}$ are known. 

As shown in \cite{Bicak 80}, the effective potentials $V_{l\pm}^{\mathrm{odd,even}}$
have the same qualitative properties as the scalar-field effective
potential \cite{Bicak 80,Sela}, 
\[
F_{l}^{\mathrm{scalar}}=\frac{1}{r^{2}}\left(1-\frac{M}{r}\right)^{2}\left(L+\frac{2M}{r}-\frac{2M^{2}}{r^{2}}\right).
\]
In addition, the potentials $V_{l\pm}^{\mathrm{odd,even}}$ have the
following asymptotic behaviors near spatial infinity and near the
horizon:
\[
V_{l\pm}^{\mathrm{odd,even}}=\frac{l\left(l+1\right)}{r^{2}}+\mathcal{O}\left(\frac{M}{r^{3}}\right)\quad,\quad r\rightarrow\infty
\]
and 
\[
V_{l+}^{\mathrm{odd,even}}=\left(l+1\right)\left(l+2\right)\left(\frac{r-M}{M^{2}}\right)^{2}+\mathcal{O}\left[M^{4}\left(\frac{r-M}{M^{2}}\right)^{3}\right]\quad,\quad r\rightarrow M,
\]
\[
V_{l-}^{\mathrm{odd,even}}=\left(l-1\right)l\left(\frac{r-M}{M^{2}}\right)^{2}+\mathcal{O}\left[M^{4}\left(\frac{r-M}{M^{2}}\right)^{3}\right]\quad,\quad r\rightarrow M.
\]
With the help of Eqs. \eqref{eq: r(s) inf} and \eqref{eq: r(s) horizon},
we can write these asymptotic behaviors in terms of $r_{*}$, 
\begin{equation}
V_{l\pm}^{\mathrm{odd,even}}=\frac{l\left(l+1\right)}{r_{*}^{2}}+4M\frac{l\left(l+1\right)}{\left|r_{*}\right|^{3}}\ln\left(\left|r_{*}\right|/M\right)+\mathcal{O}\left(Mr_{*}^{-3}\right)\quad,\quad r_{*}\rightarrow\infty\label{eq: V asymp inf}
\end{equation}
and
\begin{equation}
V_{l+}^{\mathrm{odd,even}}=\frac{\left(l+1\right)\left(l+2\right)}{r_{*}^{2}}+4M\frac{\left(l+1\right)\left(l+2\right)}{\left|r_{*}\right|^{3}}\ln\left(\left|r_{*}\right|/M\right)+\mathcal{O}\left(Mr_{*}^{-3}\right)\quad,\quad r_{*}\rightarrow-\infty,\label{eq: Vp asymp hor}
\end{equation}
\begin{equation}
V_{l-}^{\mathrm{odd,even}}=\frac{\left(l-1\right)l}{r_{*}^{2}}+4M\frac{\left(l-1\right)l}{\left|r_{*}\right|^{3}}\ln\left(\left|r_{*}\right|/M\right)+\mathcal{O}\left(Mr_{*}^{-3}\right)\quad,\quad r_{*}\rightarrow-\infty.\label{eq: Vm asymp hor}
\end{equation}
In these asymptotic regions, the scalar-field potential $F_{l}^{\mathrm{scalar}}$
takes the asymptotic forms 
\[
F_{l}^{\mathrm{scalar}}=\frac{l\left(l+1\right)}{r^{2}}+\mathcal{O}\left(\frac{M}{r^{3}}\right)\quad,\quad r\rightarrow\infty
\]
and 
\[
F_{l}^{\mathrm{scalar}}=l\left(l+1\right)\left(\frac{r-M}{M^{2}}\right)^{2}+\mathcal{O}\left[M^{4}\left(\frac{r-M}{M^{2}}\right)^{3}\right]\quad,\quad r\rightarrow M.
\]
Using Eqs. \eqref{eq: r(s) inf} and \eqref{eq: r(s) horizon} as
before, we can write these in terms of $r_{*}$,
\begin{equation}
F_{l}^{\mathrm{scalar}}=\frac{l\left(l+1\right)}{r_{*}^{2}}+4M\frac{l\left(l+1\right)}{\left|r_{*}\right|^{3}}\ln\left(\left|r_{*}\right|/M\right)+\mathcal{O}\left(Mr_{*}^{-3}\right)\quad,\quad r_{*}\rightarrow\pm\infty.\label{eq: F asymp}
\end{equation}
It is now clear from Eqs. \eqref{eq: V asymp inf} and \eqref{eq: F asymp}
that the leading and the next-to-leading (in $r_{*}^{-1}$) order
terms in the potentials $F_{l}^{\mathrm{scalar}}$, $V_{l+}^{\mathrm{odd,even}}$,
and $V_{l-}^{\mathrm{odd,even}}$ at the limit $r_{*}\rightarrow\infty$
are the same. That is, the centrifugal potential term and the leading
curvature-induced term that appear in these potentials at the limit
$r_{*}\rightarrow\infty$ are the same. Therefore, at spatial infinity,
these potentials satisfy {[}to leading and next-to-leading (in $r_{*}^{-1}$)
order{]}
\begin{equation}
V_{l\pm}^{\mathrm{odd,even}}=F_{l}^{\mathrm{scalar}}\quad,\quad r_{*}\rightarrow+\infty.\label{eq: V F inf}
\end{equation}

Analogously, from Eqs. \eqref{eq: Vp asymp hor}, \eqref{eq: Vm asymp hor},
and \eqref{eq: F asymp}, one can readily see that near the horizon
($r_{*}\rightarrow-\infty$), these potentials satisfy {[}to leading
and next-to-leading (in $r_{*}^{-1}$) order{]}
\begin{equation}
V_{l\pm}^{\mathrm{odd,even}}=F_{l\pm1}^{\mathrm{scalar}}\quad,\quad r_{*}\rightarrow-\infty.\label{eq: V F hor}
\end{equation}
The reason for keeping the next-to-leading order terms (curvature-induced
terms) in the asymptotic expansions of the potentials is that these
terms are essential for determining the late-time decay of perturbations
with initial data that is of compact support and is well separated
from the horizon (see below for more details). 

As shown and discussed in Refs. \cite{Bicak 80,Lucietti}, for regular
initial data, we get that $\Psi_{\pm}$ are regular at the horizon
and at future null infinity (FNI). 

Therefore, $\Psi_{\pm}$ satisfy wave equations with effective potentials
that have the same asymptotic forms as the scalar-field potential
and have regular boundary conditions. Now, since these properties
are the only ones we needed in \cite{Sela} in order to determine
the late-time decay of the scalar perturbations, we can use our results
and experience from \cite{Sela} to determine the late-time decay
of $\Psi_{\pm}$. Then, as a result, we get the late-time behavior
of the coupled perturbations.

\section{initial-value setup}

When analyzing the dynamics (and, in particular, the late-time decay)
of perturbations, a key ingredient is the specification of initial-value
data. We consider characteristic initial-value problem for $\Psi_{\pm}$,
just like in \cite{Sela}, for which the initial value of the perturbations
is specified along two intersecting radial null rays, $u=\mathrm{const}$
and $v=\mathrm{const}$, where $u$ and $v$ are the usual null coordinates,
$u=t-r_{*}$ and $v=t+r_{*}$.

According to Refs. \cite{Lucietti,Aretakis,Sela}, the various forms
of late-time decay of \emph{scalar} perturbations of an ERN black
hole can be classified according to the values of the Aretakis and
Newman-Penrose (NP) constants associated with their initial-value
data. If either the Aretakis constant or the NP constant is nonzero,
the late-time decay would be $\sim t^{-\left(2l+2\right)}$; otherwise,
it would be $\sim t^{-\left(2l+3\right)}$. Now, since we have (for
$l\geq2$) four types of gauge invariant combinations $\Psi_{\pm}$
(two combinations $\pm$ for each parity) that behave as scalar fields
(at least as long as their late-time decay is concerned) and four
types of initial-value data for scalar perturbations (according to
whether their Aretakis and NP constants vanish), we have a total of
$4^{4}=256$ scenarios.%
\footnote{For $l=1$, only $\Psi_{+}$ is meaningful and therefore we have $4^{2}=16$
scenarios.%
} Since this number is very large, we shall only consider the four
scenarios that make no difference between the initial data of the
various combinations and parities. In other words, we consider all
the different $\Psi_{\pm}$ on an equal footing with respect to the
type of initial data. The four scenarios result from the four types
of initial-value data, defined as follows.
\begin{description}
\item [{Type~A.}] ``Horizon-based initial data'' -- Initial data for
which $\Psi_{\pm}$ have nonvanishing Aretakis constants and vanishing
NP constants. That is, we consider initial data with generic regular
behavior across the horizon.
\item [{Type~B.}] ``FNI-based initial data'' -- Initial data for which
$\Psi_{\pm}$ have vanishing Aretakis constants and nonvanishing NP
constants.
\item [{Type~C.}] Initial data for which $\Psi_{\pm}$ have nonvanishing
Aretakis constants and nonvanishing NP constants. This type is essentially
a combination of the two types A and B.
\item [{Type~D.}] Initial data for which $\Psi_{\pm}$ have vanishing
Aretakis and NP constants.
\end{description}
Note that since the asymptotic form of the effective potentials $V_{l\pm}^{\mathrm{odd,even}}$
is different in the two limits $r_{*}\rightarrow\infty$ and $r_{*}\rightarrow-\infty$
{[}cf. Eqs. \eqref{eq: F asymp}, \eqref{eq: V F inf}, and \eqref{eq: V F hor}{]},
the late-time decays of $\Psi_{\pm}$ that result from the two types
of initial data A and B are generally different (as opposed to scalar
perturbations, where the decay is the same for all three types of
initial data A,B, and C). 

Note also that for initial data of types A and C, $\Psi_{\pm}$ that
correspond to a certain $l$ value have nonvanishing $l\pm1$ Aretakis
constants.

We can now turn to calculate the late-time tails of $\Psi_{\pm}$
that correspond to the four types of initial-value data A-D.

\section{late-time tails of $\Psi_{\pm}$}

As discussed above, we can determine the late-time decay of $\Psi_{\pm}$
using our understanding of scalar-field perturbations. As mentioned,
in the case of scalar perturbations of an ERN black hole, if either
the Aretakis constant or the NP constant is nonzero, the late-time
decay would be $\sim t^{-\left(2l+2\right)}$; otherwise, it would
be $\sim t^{-\left(2l+3\right)}$. As discussed in detail in \cite{Sela},
the $t^{-\left(2l+2\right)}$ decay is associated with the centrifugal
potential term that appears in the asymptotic form of the scalar-field
effective potential {[}cf. Eq. \eqref{eq: F asymp}{]}. Moreover,
it is well-known that for initial data with vanishing Aretakis and
NP constants, the leading tail $t^{-\left(2l+3\right)}$ (for scalar
perturbations) is formed from the scattering of the perturbations
off the leading, curvature-induced part of the effective potential
{[}see Eq. \eqref{eq: F asymp} for example of this part of the effective
potential{]}. See Refs. {[}17-19,12-14{]} for further details. Note
that this tail is also formed if the initial data have a generic regular
behavior across the horizon or FNI (corresponds to types A, B, and
C); However, in the scalar case, the tails that result from the centrifugal
part of the potential {[}$\sim t^{-\left(2l+2\right)}${]} dominates
these curvature-induced tails {[}$\sim t^{-\left(2l+3\right)}${]}.
We will see below that this is not always the case for $\Psi_{\pm}$. 

Now, since Eqs. \eqref{eq: V F hor} and \eqref{eq: V F inf} apply
to both parts (centrifugal and leading, curvature induced) of the
asymptotic effective potential, we can determine the late-time decay
of $\Psi_{\pm}$ (that correspond to certain initial data) by considering
the two types of contributions coming from the two asymptotic regions
$r_{*}\rightarrow+\infty$ and $r_{*}\rightarrow-\infty$. Specifically,
for each asymptotic region, we begin by considering the two types
of contributions (associated with the centrifugal and curvature-induced
parts of the potential) to the late-time tails of \emph{scalar-field}
perturbations with the same kind of initial data; then, the corresponding
contributions to the tails of $\Psi_{\pm}$ are determined according
to Eqs. \eqref{eq: V F hor} and \eqref{eq: V F inf}: The contributions
to the tails of $\Psi_{\pm}$ from the region $r_{*}\rightarrow-\infty$
are the same as the tails of the scalar perturbations, but with a
different $l$ value: $l\rightarrow l\pm1$; and the contributions
to the tails of $\Psi_{\pm}$ from the region $r_{*}\rightarrow+\infty$
are the same as the tails of the scalar perturbations (with the same
$l$ value).

Now, after discussing the reasoning, we find the late-time decay of
$\Psi_{\pm}$ for the initial data A-D.

\subsection{Type A initial data}

Since the NP constants (built from $\Psi_{\pm}$) vanish, the only
contribution from the region $r_{*}\rightarrow+\infty$ comes from
the leading, curvature-induced part of the potential, and therefore,
results in a tail $t^{-\left(2l+3\right)}$ for both scalar perturbations
and $\Psi_{\pm}$, in accordance with Eq. \eqref{eq: V F inf}. In
the region $r_{*}\rightarrow-\infty$, nonvanishing Aretakis constants,
associated with type A initial data, yield the leading \emph{scalar-field}
tail $t^{-\left(2l+2\right)}$ (that results from the centrifugal
part of the potential). The corresponding contributions to the tails
of $\Psi_{\pm}$ are $t^{-\left(2l+4\right)}$ for $\Psi_{+}$ ($l\rightarrow l+1$)
and $t^{-2l}$ for $\Psi_{-}$ ($l\rightarrow l-1$), in accordance
with Eq. \eqref{eq: V F hor}. 

In summary, the leading tail of $\Psi_{+}$ is $t^{-\left(2l+3\right)}$
(associated with the region $r_{*}\rightarrow+\infty$) and the leading
tail of $\Psi_{-}$ is $t^{-2l}$ (associated with the region $r_{*}\rightarrow-\infty$).
Here, in the case of $\Psi_{+}$, we see an example where a curvature-induced
tail can dominate a tail that results from the centrifugal (``flat
space'') part of the potential. 

We shall use superscripts to denote the type of initial data that
corresponds to the quantity under consideration. We get 
\[
\Psi_{+}^{\left(A\right)}\sim t^{-\left(2l+3\right)}\quad,\quad\Psi_{-}^{\left(A\right)}\sim t^{-2l}.
\]

\subsection{Type B initial data}

In this case, it is clear that the contribution from the region $r_{*}\rightarrow-\infty$
to the \emph{scalar-field} tail is $t^{-\left(2l+3\right)}$. As a
result, the contributions to $\Psi_{\pm}$ are $t^{-\left(2l+5\right)}$
for $\Psi_{+}$ and $t^{-\left(2l+1\right)}$ for $\Psi_{-}$. 

The leading contribution from the region $r_{*}\rightarrow\infty$
to the \emph{scalar-field} tail is $t^{-\left(2l+2\right)}$, which
is also the tail of both $\Psi_{\pm}$ by virtue of Eq. \eqref{eq: V F inf}. 

In summary, we get the leading tails 
\[
\Psi_{+}^{\left(B\right)}\sim t^{-\left(2l+2\right)}\quad,\quad\Psi_{-}^{\left(B\right)}\sim t^{-\left(2l+1\right)}.
\]

\subsection{Type C initial data}

In this case, both regions ($r_{*}\rightarrow\pm\infty$) contribute
to the \emph{scalar-field} tail $t^{-\left(2l+2\right)}$ (this is
the leading tail). The contribution to $\Psi_{\pm}$ from the region
$r_{*}\rightarrow\infty$ is of course the same. The contribution
to $\Psi_{\pm}$ from the region $r_{*}\rightarrow-\infty$ is $t^{-\left(2l+4\right)}$
for $\Psi_{+}$ and $t^{-2l}$ for $\Psi_{-}$. 

In summary, we get the leading tails 
\[
\Psi_{+}^{\left(C\right)}\sim t^{-\left(2l+2\right)}\quad,\quad\Psi_{-}^{\left(C\right)}\sim t^{-2l}.
\]

\subsection{Type D initial data}

Both regions ($r_{*}\rightarrow\pm\infty$) contribute to the \emph{scalar-field}
tail $t^{-\left(2l+3\right)}$. The contribution to $\Psi_{\pm}$
from the region $r_{*}\rightarrow\infty$ is of course the same. The
contribution to $\Psi_{\pm}$ from the region $r_{*}\rightarrow-\infty$
is $t^{-\left(2l+5\right)}$ for $\Psi_{+}$ and $t^{-\left(2l+1\right)}$
for $\Psi_{-}$. 

In summary, we get the leading tails 
\[
\Psi_{+}^{\left(D\right)}\sim t^{-\left(2l+3\right)}\quad,\quad\Psi_{-}^{\left(D\right)}\sim t^{-\left(2l+1\right)}.
\]

\section{late-time decay of moncrief's gauge invariant quantities}

Now, after we found the late-time decay of $\Psi_{\pm}$ for initial
data A-D, we can find the corresponding late-time tails of Moncrief's
gauge invariant quantities. We begin by writing Moncrief's quantities
in terms of $\Psi_{\pm}$ (for $l\geq2$) by solving Eqs. \eqref{P+-}
and \eqref{R+-} for them. For odd parity perturbations, we get ($l\geq2$)
\[
\hat{\pi}_{g}=\left(2\sigma\right)^{-1/2}\left[\left(\sigma-3M\right)^{1/2}P_{+}+\left(\sigma+3M\right)^{1/2}P_{-}\right]
\]
and
\[
\hat{\pi}_{f}=\left(2\sigma\right)^{-1/2}\left[\left(\sigma+3M\right)^{1/2}P_{+}-\left(\sigma-3M\right)^{1/2}P_{-}\right],
\]
and for even parity perturbations, we get ($l\geq2$)
\[
H=\left(2\sigma\right)^{-1/2}\left[\left(\sigma+3M\right)^{1/2}R_{+}+\left(\sigma-3M\right)^{1/2}R_{-}\right]
\]
and
\[
Q=\left(2\sigma\right)^{-1/2}\left[-\left(\sigma-3M\right)^{1/2}R_{+}+\left(\sigma+3M\right)^{1/2}R_{-}\right].
\]
Note that $P_{\pm}$ and $R_{\pm}$ are just $\Psi_{\pm}$ for odd
and even perturbations, respectively. Note that for $l\geq2$, $\sigma\geq5M$
{[}see Eq. \eqref{eq: sig}{]}.

For $l=1$, $\Psi_{+}$ denotes $P_{f}$ and $H$ for odd and even
perturbations, respectively ($\Psi_{-}$ has no meaning for $l=1$). 

Now, we can readily obtain the late-time decay of Moncrief's quantities
by direct substitution of the results from the previous section, and
identification of the leading tail.

\subsection{Type A initial data}

\subsubsection{Odd and even parity perturbations with $l\geq2$}

Between the two quantities $\Psi_{+}^{\left(A\right)}$ and $\Psi_{-}^{\left(A\right)}$,
the one that decays faster is $\Psi_{+}^{\left(A\right)}$. Therefore,
the decay is the same as that of $\Psi_{-}^{\left(A\right)}$, and
we get 

\[
\hat{\pi}_{g}^{\left(A\right)},\hat{\pi}_{f}^{\left(A\right)},H^{\left(A\right)},Q^{\left(A\right)}\sim t^{-2l}.
\]

\subsubsection{Odd and even parity perturbations with $l=1$}

In these two cases, the decay is simply the same as that of $\Psi_{+}^{\left(A\right)}$
with $l=1$. Therefore, 
\[
P_{f}^{\left(A\right)},H^{\left(A\right)}\sim t^{-5}.
\]

\subsection{Type B initial data}

\subsubsection{Odd and even parity perturbations with $l\geq2$}

Between $\Psi_{+}^{\left(B\right)}$ and $\Psi_{-}^{\left(B\right)}$,
the one that decays faster is $\Psi_{+}^{\left(B\right)}$. Therefore,
the decay is the same as that of $\Psi_{-}^{\left(B\right)}$, and
we get 

\[
\hat{\pi}_{g}^{\left(B\right)},\hat{\pi}_{f}^{\left(B\right)},H^{\left(B\right)},Q^{\left(B\right)}\sim t^{-\left(2l+1\right)}.
\]

\subsubsection{Odd and even parity perturbations with $l=1$}

The decay is the same as that of $\Psi_{+}^{\left(B\right)}$ with
$l=1$. Therefore, 
\[
P_{f}^{\left(B\right)},H^{\left(B\right)}\sim t^{-4}.
\]

\subsection{Type C initial data}

\subsubsection{Odd and even parity perturbations with $l\geq2$}

Between $\Psi_{+}^{\left(C\right)}$ and $\Psi_{-}^{\left(C\right)}$,
the one that decays faster is $\Psi_{+}^{\left(C\right)}$. Therefore,
the decay is the same as that of $\Psi_{-}^{\left(C\right)}$, and
we get 

\[
\hat{\pi}_{g}^{\left(C\right)},\hat{\pi}_{f}^{\left(C\right)},H^{\left(C\right)},Q^{\left(C\right)}\sim t^{-2l}.
\]

\subsubsection{Odd and even parity perturbations with $l=1$}

The decay is the same as that of $\Psi_{+}^{\left(C\right)}$ with
$l=1$. Therefore, 
\[
P_{f}^{\left(C\right)},H^{\left(C\right)}\sim t^{-4}.
\]

\subsection{Type D initial data}

\subsubsection{Odd and even parity perturbations with $l\geq2$}

Between $\Psi_{+}^{\left(D\right)}$ and $\Psi_{-}^{\left(D\right)}$,
the one that decays faster is $\Psi_{+}^{\left(D\right)}$. Therefore,
the decay is the same as that of $\Psi_{-}^{\left(D\right)}$, and
we get 

\[
\hat{\pi}_{g}^{\left(D\right)},\hat{\pi}_{f}^{\left(D\right)},H^{\left(D\right)},Q^{\left(D\right)}\sim t^{-\left(2l+1\right)}.
\]

\subsubsection{Odd and even parity perturbations with $l=1$}

The decay is the same as that of $\Psi_{+}^{\left(D\right)}$ with
$l=1$. Therefore, 
\[
P_{f}^{\left(D\right)},H^{\left(D\right)}\sim t^{-5}.
\]

\section{late-time decay of the perturbations of the metric tensor and the
electromagnetic field tensor}

In order to find the late-time tails of coupled perturbations, we
employ the results of Ref. \cite{Bicak 79}, where (in Sec. 3) the
components of the metric tensor and the electromagnetic field tensor
were given in terms of Moncrief's gauge invariant quantities. The
gauge used in these expressions is the Regge-Wheeler gauge (for $l\geq2$).
For $l=1$ perturbations, specific gauges that simplify the expressions
further were chosen such that, in addition to the metric perturbation
components that vanish for $l\geq2$ due to the gauge choice, we have
$\delta g_{r\phi}=0$ for odd parity and $\delta g_{\theta\theta}=\delta g_{\phi\phi}=0$
for even parity.

It is important to note that $l=1$ odd perturbations include stationary
perturbations that correspond to a slowly rotating Kerr-Newman black
hole. Specifically, $l=1$ odd perturbations with $P_{f}=0$ (in ERN
spacetime) generally yield {[}see Eqs. (61) and (62) in \cite{Bicak 79}{]}
\[
\delta g_{t\phi}=-\left(\frac{2M}{r}-\frac{M^{2}}{r^{2}}\right)\sin^{2}\left(\theta\right)\delta a\quad,\quad\delta A_{\phi}=\pm\frac{2M}{r}\sin^{2}\left(\theta\right)\delta a,
\]
where $\delta a$ is a small constant determined by the initial data
(and corresponds to the rotation parameter of the black hole), $A_{\mu}$
is the electromagnetic potential, and the $\pm$ sign corresponds
to the charge of the black hole ($\pm M$). In this paper, we ignore
this kind of perturbations; by $l=1$ odd perturbations, we mean perturbations
beyond the stationary Kerr-Newman ones or perturbations with $\delta a=0$.
These perturbations decay at late time and have a tail. 

After substituting the results from the previous section into the
expressions from \cite{Bicak 79} and identifying the leading tail,
we get the late-time decay of the coupled perturbations. The results
are presented below (all the components that are not written down
are either obtained by symmetry or vanish).

\subsection{Type A initial data}

\subsubsection{Odd parity perturbations with $l\geq2$}

\[
\delta g_{t\phi}^{\left(A\right)}\sim t^{-2l}\quad,\quad\delta g_{r\phi}^{\left(A\right)}\sim t^{-\left(2l+1\right)},
\]
\[
\delta F_{t\phi}^{\left(A\right)}\sim t^{-\left(2l+1\right)}\quad,\quad\delta F_{r\phi}^{\left(A\right)}\sim t^{-2l}\quad,\quad\delta F_{\theta\phi}^{\left(A\right)}\sim t^{-2l}.
\]

\subsubsection{Odd parity perturbations with $l=1$}

\[
\delta g_{t\phi,l=1}^{\left(A\right)}\sim t^{-5}
\]
\[
\delta F_{t\phi,l=1}^{\left(A\right)}\sim t^{-6}\quad,\quad\delta F_{r\phi,l=1}^{\left(A\right)}\sim t^{-5}\quad,\quad\delta F_{\theta\phi,l=1}^{\left(A\right)}\sim t^{-5}.
\]

\subsubsection{Even parity perturbations with $l\geq2$}

\[
\delta g_{tt}^{\left(A\right)}\sim t^{-2l}\quad,\quad\delta g_{rr}^{\left(A\right)}\sim t^{-2l}\quad,\quad\delta g_{\theta\theta}^{\left(A\right)}\sim t^{-2l}\quad,\quad\delta g_{\phi\phi}^{\left(A\right)}\sim t^{-2l}\quad,\quad\delta g_{rt}^{\left(A\right)}\sim t^{-\left(2l+1\right)},
\]
\[
\delta F_{tr}^{\left(A\right)}\sim t^{-2l}\quad,\quad\delta F_{t\theta}^{\left(A\right)}\sim t^{-2l}\quad,\quad\delta F_{r\theta}^{\left(A\right)}\sim t^{-\left(2l+1\right)}.
\]

\subsubsection{Even parity perturbations with $l=1$}

\[
\delta g_{tt}^{\left(A\right)}\sim t^{-5}\quad,\quad\delta g_{rr}^{\left(A\right)}\sim t^{-5}\quad,\quad\delta g_{rt}^{\left(A\right)}\sim t^{-6},
\]
\[
\delta F_{tr}^{\left(A\right)}\sim t^{-5}\quad,\quad\delta F_{t\theta}^{\left(A\right)}\sim t^{-5}\quad,\quad\delta F_{r\theta}^{\left(A\right)}\sim t^{-6}.
\]

\subsection{Type B initial data}

\subsubsection{Odd parity perturbations with $l\geq2$}

\[
\delta g_{t\phi}^{\left(B\right)}\sim t^{-\left(2l+1\right)}\quad,\quad\delta g_{r\phi}^{\left(B\right)}\sim t^{-\left(2l+2\right)},
\]
\[
\delta F_{t\phi}^{\left(B\right)}\sim t^{-\left(2l+2\right)}\quad,\quad\delta F_{r\phi}^{\left(B\right)}\sim t^{-\left(2l+1\right)}\quad,\quad\delta F_{\theta\phi}^{\left(B\right)}\sim t^{-\left(2l+1\right)}.
\]

\subsubsection{Odd parity perturbations with $l=1$}

\[
\delta g_{t\phi,l=1}^{\left(B\right)}\sim t^{-4}
\]
\[
\delta F_{t\phi,l=1}^{\left(B\right)}\sim t^{-5}\quad,\quad\delta F_{r\phi,l=1}^{\left(B\right)}\sim t^{-4}\quad,\quad\delta F_{\theta\phi,l=1}^{\left(B\right)}\sim t^{-4}.
\]

\subsubsection{Even parity perturbations with $l\geq2$}

\[
\delta g_{tt}^{\left(B\right)}\sim t^{-\left(2l+1\right)}\quad,\quad\delta g_{rr}^{\left(B\right)}\sim t^{-\left(2l+1\right)}\quad,\quad\delta g_{\theta\theta}^{\left(B\right)}\sim t^{-\left(2l+1\right)}\quad,\quad\delta g_{\phi\phi}^{\left(B\right)}\sim t^{-\left(2l+1\right)}\quad,\quad\delta g_{rt}^{\left(B\right)}\sim t^{-\left(2l+2\right)},
\]
\[
\delta F_{tr}^{\left(B\right)}\sim t^{-\left(2l+1\right)}\quad,\quad\delta F_{t\theta}^{\left(B\right)}\sim t^{-\left(2l+1\right)}\quad,\quad\delta F_{r\theta}^{\left(B\right)}\sim t^{-\left(2l+2\right)}.
\]

\subsubsection{Even parity perturbations with $l=1$}

\[
\delta g_{tt}^{\left(B\right)}\sim t^{-4}\quad,\quad\delta g_{rr}^{\left(B\right)}\sim t^{-4}\quad,\quad\delta g_{rt}^{\left(B\right)}\sim t^{-5},
\]
\[
\delta F_{tr}^{\left(B\right)}\sim t^{-4}\quad,\quad\delta F_{t\theta}^{\left(B\right)}\sim t^{-4}\quad,\quad\delta F_{r\theta}^{\left(B\right)}\sim t^{-5}.
\]

\subsection{Type C initial data}

\subsubsection{Odd parity perturbations with $l\geq2$}

\[
\delta g_{t\phi}^{\left(C\right)}\sim t^{-2l}\quad,\quad\delta g_{r\phi}^{\left(C\right)}\sim t^{-\left(2l+1\right)},
\]
\[
\delta F_{t\phi}^{\left(C\right)}\sim t^{-\left(2l+1\right)}\quad,\quad\delta F_{r\phi}^{\left(C\right)}\sim t^{-2l}\quad,\quad\delta F_{\theta\phi}^{\left(C\right)}\sim t^{-2l}.
\]

\subsubsection{Odd parity perturbations with $l=1$}

\[
\delta g_{t\phi,l=1}^{\left(C\right)}\sim t^{-4}
\]
\[
\delta F_{t\phi,l=1}^{\left(C\right)}\sim t^{-5}\quad,\quad\delta F_{r\phi,l=1}^{\left(C\right)}\sim t^{-4}\quad,\quad\delta F_{\theta\phi,l=1}^{\left(C\right)}\sim t^{-4}.
\]

\subsubsection{Even parity perturbations with $l\geq2$}

\[
\delta g_{tt}^{\left(C\right)}\sim t^{-2l}\quad,\quad\delta g_{rr}^{\left(C\right)}\sim t^{-2l}\quad,\quad\delta g_{\theta\theta}^{\left(C\right)}\sim t^{-2l}\quad,\quad\delta g_{\phi\phi}^{\left(C\right)}\sim t^{-2l}\quad,\quad\delta g_{rt}^{\left(C\right)}\sim t^{-\left(2l+1\right)},
\]
\[
\delta F_{tr}^{\left(C\right)}\sim t^{-2l}\quad,\quad\delta F_{t\theta}^{\left(C\right)}\sim t^{-2l}\quad,\quad\delta F_{r\theta}^{\left(C\right)}\sim t^{-\left(2l+1\right)}.
\]

\subsubsection{Even parity perturbations with $l=1$}

\[
\delta g_{tt}^{\left(C\right)}\sim t^{-4}\quad,\quad\delta g_{rr}^{\left(C\right)}\sim t^{-4}\quad,\quad\delta g_{rt}^{\left(C\right)}\sim t^{-5},
\]
\[
\delta F_{tr}^{\left(C\right)}\sim t^{-4}\quad,\quad\delta F_{t\theta}^{\left(C\right)}\sim t^{-4}\quad,\quad\delta F_{r\theta}^{\left(C\right)}\sim t^{-5}.
\]

\subsection{Type D initial data}

\subsubsection{Odd parity perturbations with $l\geq2$}

\[
\delta g_{t\phi}^{\left(D\right)}\sim t^{-\left(2l+1\right)}\quad,\quad\delta g_{r\phi}^{\left(D\right)}\sim t^{-\left(2l+2\right)},
\]
\[
\delta F_{t\phi}^{\left(D\right)}\sim t^{-\left(2l+2\right)}\quad,\quad\delta F_{r\phi}^{\left(D\right)}\sim t^{-\left(2l+1\right)}\quad,\quad\delta F_{\theta\phi}^{\left(D\right)}\sim t^{-\left(2l+1\right)}.
\]

\subsubsection{Odd parity perturbations with $l=1$}

\[
\delta g_{t\phi,l=1}^{\left(D\right)}\sim t^{-5}
\]
\[
\delta F_{t\phi,l=1}^{\left(D\right)}\sim t^{-6}\quad,\quad\delta F_{r\phi,l=1}^{\left(D\right)}\sim t^{-5}\quad,\quad\delta F_{\theta\phi,l=1}^{\left(D\right)}\sim t^{-5}.
\]

\subsubsection{Even parity perturbations with $l\geq2$}

\[
\delta g_{tt}^{\left(D\right)}\sim t^{-\left(2l+1\right)}\quad,\quad\delta g_{rr}^{\left(D\right)}\sim t^{-\left(2l+1\right)}\quad,\quad\delta g_{\theta\theta}^{\left(D\right)}\sim t^{-\left(2l+1\right)}\quad,\quad\delta g_{\phi\phi}^{\left(D\right)}\sim t^{-\left(2l+1\right)}\quad,\quad\delta g_{rt}^{\left(D\right)}\sim t^{-\left(2l+2\right)},
\]
\[
\delta F_{tr}^{\left(D\right)}\sim t^{-\left(2l+1\right)}\quad,\quad\delta F_{t\theta}^{\left(D\right)}\sim t^{-\left(2l+1\right)}\quad,\quad\delta F_{r\theta}^{\left(D\right)}\sim t^{-\left(2l+2\right)}.
\]

\subsubsection{Even parity perturbations with $l=1$}

\[
\delta g_{tt}^{\left(D\right)}\sim t^{-5}\quad,\quad\delta g_{rr}^{\left(D\right)}\sim t^{-5}\quad,\quad\delta g_{rt}^{\left(D\right)}\sim t^{-6},
\]
\[
\delta F_{tr}^{\left(D\right)}\sim t^{-5}\quad,\quad\delta F_{t\theta}^{\left(D\right)}\sim t^{-5}\quad,\quad\delta F_{r\theta}^{\left(D\right)}\sim t^{-6}.
\]

\section{concluding remarks}

In this paper, we employed the results of \cite{Sela} and \cite{Bicak 79}
and found the late-time decay of coupled electromagnetic and gravitational
perturbations outside an extremal charged black hole. In particular,
we have explicitly shown that the coupled perturbations do decay (except
$l=1$ odd perturbations that might correspond to a slowly rotating
Kerr-Newman black hole) and found the decay rate in a way that is
consistent with Refs. \cite{Burko,Lucietti,Sela}. In addition , we
can notice some nontrivial features of the decay of coupled perturbations
that do not appear in the scalar case. For example, we can easily
see from the explicit formulas for the decay rates of the perturbations
of the metric tensor and the electromagnetic field tensor (from the
previous section) that for type A initial data, the quadrupole ($l=2$)
perturbations generally decay more slowly than the dipole ($l=1$)
perturbations, in contrast to the corresponding decays of scalar perturbations.%
\footnote{In \cite{Bicak 80}, Bicak pointed into a similar observation. However,
it was based on apparently wrong results for the late-time decay of
scalar perturbations.%
} For type B initial data, we get the opposite behavior, and quadrupole
perturbations generally decay faster. 

Coupled electromagnetic and gravitational perturbations were also
investigated in \cite{Lucietti} in the context of the horizon instability
of an ERN black hole. In \cite{Lucietti}, it was shown that if the
coupled perturbations and their derivatives decay outside the horizon
(more specifically, that $\Psi_{\pm}$ and their derivatives decay),
then a certain linear combination of $\Psi_{\pm}$ and its $r$ derivatives
blows up at late time on the horizon.%
\footnote{This analysis extends the one performed by Aretakis \cite{Aretakis}
(for scalar perturbations) to coupled perturbations.%
} Since we have shown that such decay of $\Psi_{\pm}$ takes place,
we may say that an instability of an ERN black hole occurs for coupled
(linearized) gravitational and electromagnetic perturbations. 

It would be interesting, as a future research, to try to find the
full leading late-time behavior of the coupled perturbations. In order
to do it, one can try to employ the so-called ``late-time expansion'',
presented, for example, in Refs. \cite{Amos,Barack b}, and use the
exact stationary solutions given in \cite{Bicak 77}. If there are
nonvanishing Aretakis or NP constants (for $\Psi_{\pm}$), one can
also employ them for the calculation.

An additional natural extension of the current research and the one
performed in \cite{Sela} would be the study of Yang-Mills fields
on the exterior of the ERN black hole. Such a study, for example,
was carried out in \cite{Bizon =000026 Kahl} for the particular case
of a spherically symmetric $SU\left(2\right)$ Yang-Mills field. It
would be interesting to investigate it further and check whether the
obtained results are related to those of \cite{Sela} and the present
paper.

\section*{ACKNOWLEDGMENTS}

I would like to thank Professor A. Ori for helpful discussions.


\begin{thebibliography}{10}
\bibitem{Zerilli}F.J. Zerilli, \emph{Perturbation analysis for gravitational
and electromagnetic radiation in a Reissner-Nordstrom geometry}, Phys.
Rev. D \textbf{9}, 860 (1974).

\bibitem{Moncrief a}V. Moncrief, \emph{Odd-parity stability of a
Reissner-Nordstrom black hole}, Phys. Rev. D \textbf{9}, 2707 (1974).

\bibitem{Moncrief b}V. Moncrief, \emph{Stability of Reissner-Nordstrom
black holes}, Phys. Rev. D \textbf{10}, 1057 (1974).

\bibitem{Moncrief c}V. Moncrief, \emph{Gauge-invariant perturbations
of Reissner-Nordstrom black holes}, Phys. Rev. D \textbf{12}, 1526
(1975).

\bibitem{Chitre}D. Chitre, \emph{Perturbations of Charged Black Holes},
Phys. Rev. D \textbf{13}, 2713 (1976).

\bibitem{Lee}C.H. Lee, \emph{Coupled gravitational and electromagnetic
perturbations around a charged black hole}, J. Math. Phys. \textbf{17},
1226 (1976).

\bibitem{Lee =000026 McGlinn}C.H. Lee and W. D. McGlinn, \emph{Uniqueness
of perturbation of a Reissner-Nordstrom black hole}, J. Math. Phys.
\textbf{17}, 2159 (1976).

\bibitem{Chandrasekhar}S. Chandrasekhar, \emph{The Mathematical Theory
of Black Holes} (Oxford, UK, Clarendon, 1985).

\bibitem{Bicak 79}J. Bicak, \emph{On the theories of the interacting
perturbations of the Reissner-Nordstrom black hole}, Czech. J. Phys.
B \textbf{29}, 945 (1979).

\bibitem{Bicak 80}J. Bicak, \emph{Gravitational collapse with charge
and small asymmetries. II. Interacting electromagnetic and gravitational
perturbations}, Gen. Relativ. Gravit. \textbf{12}, 195 (1980).

\bibitem{Lucietti} J. Lucietti, K. Murata, H. S. Reall, and N. Tanahashi,
\emph{On the horizon instability of an extreme Reissner-Nordstrom
black hole}, J. High Energy Phys. 03 (2013) 035.

\bibitem{Bicak 72}J. Bicak, \emph{Gravitational collapse with charge
and small asymmetries. I. Scalar perturbations}, Gen. Relativ. Gravit.
\textbf{3}, 331 (1972).

\bibitem{Sela}O. Sela, \emph{Late-time decay of perturbations outside
extremal charged black hole}, Phys. Rev. D \textbf{93}, 024054 (2016).

\bibitem{Price a}R. H. Price, \emph{Nonspherical perturbations of
relativistic gravitational collapse. I. Scalar and gravitational perturbations},
Phys. Rev. D \textbf{5}, 2419 (1972).

\bibitem{Price b}R. H. Price, \emph{Nonspherical perturbations of
relativistic gravitational collapse. II. Integer-spin, zero-rest-mass
fields}, Phys. Rev. D \textbf{5}, 2439 (1972).

\bibitem{Aretakis}S. Aretakis, \emph{Stability and instability of
extreme Reissner-Nordstrom black hole spacetimes for linear scalar
perturbations II}, Ann. Henri Poincare \textbf{12}, 1491 (2011) and
additional references therein.

\bibitem{Burko}C. J. Blaksley and L. M. Burko, \emph{The late-time
tails in the Reissner-Nordstrom spacetime revisited}, Phys. Rev. D
\textbf{76}, 104035 (2007).

\bibitem{Barack a}L. Barack, \emph{Late time dynamics of scalar perturbations
outside black holes. I. A shell toy-model}, Phys. Rev. D \textbf{59},
044016 (1999).

\bibitem{Barack b}L. Barack, \emph{Late time dynamics of scalar perturbations
outside black holes. II. Schwarzschild geometry}, Phys. Rev. D \textbf{59},
044017 (1999).

\bibitem{Amos}A. Ori, \emph{Transmission and reflection coefficients
for a scalar field inside a charged black hole}, Phys. Rev. D \textbf{57},
2621 (1998).

\bibitem{Bicak 77}J. Bicak, \emph{Stationary interacting fields around
an extreme Reissner-Nordstrom black hole}, Phys. Lett \textbf{64A},
279 (1977).

\bibitem{Bizon =000026 Kahl}P. Bizon and M. Kahl, \emph{A Yang-Mills
field on the extremal Reissner-Nordstrom black hole}, Classical Quantum
Gravity \textbf{33}, 175013 (2016).\end{thebibliography}
\end{document}